\documentclass[aps,prl,twocolumn,showpacs]{revtex4}
\usepackage{amssymb}

\usepackage{graphicx,amsmath}
\usepackage{epstopdf}
\newcommand{\bea}{\begin{eqnarray}}
\newcommand{\eea}{\end{eqnarray}}
\newcommand{\beq}{\begin{equation}}
\newcommand{\eeq}{\end{equation}}
\newcommand{\benu}{\begin{enumerate}}
\newcommand{\enu}{\end{enumerate}}

\begin{document}
\title{Model of Quantum Criticality in He$^3$ bilayers Adsorbed on Graphite}
\date{\today}
\author{A. Benlagra and C. P\'epin}
\affiliation{CEA, DSM, Institut de Physique Théorique, IPhT, CNRS, MPPU, URA2306, Saclay, F-91191 Gif-sur-Yvette, France\\
}

\begin{abstract}
Recent experiments on He$^3$ bilayers adsorbed on graphite 
have shown striking quantum critical properties at the point where the first layer localizes.
We  model this system with the Anderson lattice plus inter-layer Coulomb repulsion in two dimensions. 
Assuming that 
quantum critical fluctuations come from a vanishing of the effective hybridization, 
we can reproduce several features of the system, including the apparent occurrence 
of two quantum critical points,
the variation of the effective mass and coherence temperature with coverage.

\end{abstract}

\maketitle
   He$^3$ layers, adsorbed on graphite have attracted substantial interest since the early eighties, 
for their remarkable properties of surface magnetism \cite{saunders, godfrin,greywall,saunders-exchange}
as well as a model system for quantum wetting transitions\cite{pricaupenko}.
 One  layer of He$^3$ atoms adsorbed  on graphite pre-plated by  HD-bilayer or  
compressed solid has been
shown to solidify at a  coverage of 4/7 of the substrate\cite{fukuyama,godfrin2,roger2}. 
This transition has been
identified as a Mott transition for the  He$^3$ fermions\cite{casey}.  
One of the leading questions in this field has been to know whether the ground state 
in the Mott phase orders magnetically or is a spin liquid\cite{fukuyama}, 
and, in the latter case, which kind of spin liquid -gapless or gapfull- it is\cite{godfrin2,ishimoto}. 
 Theoretical studies have shown that, depending on the relative strength of the  ring 
exchange parameters\cite{roger1,roger2},
 a spin liquid phase, a canted phase, or even ferromagnetic ground state 
can occur. 
Close to the Mott transition, a gapless spin liquid  ground state appears 
to be a reasonable choice\cite{gregoire}.
 Experimental studies of the solidification of the first He$^3$ layer when a 
second or a third  layer is adsorbed 
have been performed long ago \cite{greywall,godfrin3}. 
There, an enhancement of the static spin susceptibility as well
as of the specific heat coefficient was observed close to the Mott transition.
In the experiment \cite{saunders}, two layers of He$^3$ are adsorbed on Graphite pre-plated by two
layers of He$^4$.
The originality of the data \cite{saunders} lies in that it's the first system for
which, when the second He$^3$ layer arrives at promotion, the first  
layer is not yet solidified.
As a  function of the layer coverage, promotion occurs at $n_0 \simeq 6.3 nm^{-2}$ 
while the first layer's solidification occurs 
at $n_c \simeq 9.9 nm^{-2}$ 
(which corresponds roughly to a ratio of 13/19 or 
12/19 between the first He$^3$ layer and the He$^4$ substrate\cite{roger2}). 
From specific heat measurements,  the effective
 mass is seen to be enhanced like $m/m^* \sim \delta$, with $\delta=1- n/ n_c$,  
and the coherence temperature, 
below which the Fermi liquid behavior is recovered, is shown to decrease 
like $ T_{coh} \sim \delta^{1.8}$ while approaching the quantum critical point (QCP).
One striking feature of the data is that, strictly speaking, the experiment doesn't reach 
the QCP; at $n_1 \simeq 9.2 nm^{-2}$ the specific heat behavior 
shows hints of a first order transition. Moreover 
NMR studies show that the field-driven magnetization abruptly starts to grow at $n = 9.2 nm^{-2}$. 
An activation gap extracted 
from the low energy behavior of the specific heat coefficient seems
 to vanish before the quantum critical coverage is reached. 
In conclusion, the phase diagram of He$^3$ bi-layers pre-plated on two 
He$^4$ layers, seems to exhibit {\it two} mysterious phase transitions.
One at which the magnetization starts to grow, which corresponds  to 
input of a first order transition, and
one at which  the ratio $m/m^*$ and the
coherence temperature  curves extrapolate  to zero as a function of coverage.

In this letter we model the system with the Anderson lattice in two dimensions 
with the addition of inter- and intra-layer Coulomb
repulsion\cite{hewson}. Quasi-local f-fermions are identified to  the first layer He$^3$ atoms while 
light c-fermions are the second layer He$^3$ atoms. The bare hybridization corresponds to 
the hopping  between the two layers.
 In the spirit of the early studies of  bulk He$^3$ \cite{vollhardt} we 
solve this model  for infinite  Coulomb repulsion $U$ between the f-fermions, 
using a slave boson 
technique equivalent to 
Gutzwiller's variational approach. The QCP of this system is identified with
the Kondo breakdown QCP (KB-QCP) of the Anderson lattice\cite{senthil,us}; 
that is the fixed point for which
the effective hybridization vanishes and, at the same time, the f-fermions localize. 
On the disordered side of the transition,
  a spin-liquid phase is necessary to stabilize the KB-QCP.
We can reproduce the variations of the inverse effective mass and coherence temperature
 with coverage (namely $m/m^* \simeq \delta$  and $T_{coh} \simeq \delta^2$ ) in 
an intermediate temperature regime. Those variations are in good agreement with experiments, 
as seen in Fig.2 where we fit
experimental data.
 While approaching the experimental critical coverage, 
the order parameter suddenly drops, reaching the true theoretical QCP {\it before} the experimental critical coverage is reached.
We believe this sudden drop of the effective hybridization
explains the mysterious observation that the system seems to exhibit two QCPs.
 Indeed, for us the true QCP is the one at which
the hybridization goes to zero, which identifies experimentally with the point where the magnetization starts to grow at
$n= 9.2 nm^{-2}$. The experimental QCP corresponds to the extrapolation to $T=0$ of the inverse 
effective mass and coherence temperature curves obtained in the {\it intermediate energy } regime.
The drop of the order parameter obtained in our model is so abrupt that it may trigger as well
first order transitions in the low energy regime.

Our starting point is the Anderson lattice model with  inter-  and intra-layer Coulomb repulsion:
\bea \label{eqn1} 
H   & = & \sum_{
\langle i, j \rangle, \sigma} \left [  {\tilde f}^\dagger_{i \sigma}  \left (t^0_{ij} + E_0 \delta_{ij} \right ) {\tilde f}_{j \sigma}  \right  . \nonumber \\
& + & \left . 
 c^\dagger_{i \sigma} \left ( t_{ij} - \mu \delta_{ij} \right )  c_{j \sigma} \right ]   
+   V   \sum_{i \sigma}     \left   (  {\tilde f}^\dagger_{i \sigma} c_{i \sigma}  + h.c. \right ) \nonumber \\
&  + &  
\sum_i \left ( U {\tilde n}_{f, i}^2 + U_1 {\tilde n}_{f, i } n_{c, i } + U_2 n_{c,i }^2 \right )  \ , \eea 
where $(i,j)$ are the lattice sites created by the  Graphite's corrugate potential, ${\tilde f}^\dagger ( {\tilde f } )$ are the creation (annihilation) operators for the first layer's 
fermions,
$ c^\dagger (  c  )$ are the creation (annihilation) operators for the second layer's fermions, $ t_{ij} =t $ is the c-fermion's 
hopping taken as a constant, $ t^0_{ij} = \alpha t $ is the f-fermion's hopping term, $V $ is the 
hybridization corresponding to a hopping 
term between the two layers, $ E_0< 0 $ is the f-level potential and $ \mu $ is the c-fermions chemical potential. 
Physically, the localization  of the first layer is driven
by the Graphite's corrugated potential, inducing a triangular lattice in the He$^4$ bi-layers, inducing itself a 
second triangular lattice for the first He$^3$ layer,
commensurate with the substrate's one at the ``magic'' filling number 13/19\cite{roger2}.  With respect to the ``13/19'' lattice, 
the f-fermion are thought of being close to a Mott transition; the f-band is close to  half-filling and the hard core 
Coulomb repulsion leads to strongly correlated effects. We treat the effect of strong correlations by  introducing  one 
Coleman's slave boson \cite{coleman84}  which decouples the ${\tilde f}^\dagger$ creation operator at site ``i'' in the 
following way: ${\tilde f}^\dagger_{i \sigma} \rightarrow  f^\dagger_{ i \sigma}  b_i $  where  the f-spinons and 
the b-holons are subject to the constraint $ \sum_\sigma f^\dagger_{i \sigma } f_{i \sigma }  + b^\dagger_i b_i = 1$.    
The constraint is  taken into account  in a Lagrangian formulation  through a Lagrange multiplier $ \lambda $.    
 The properties of the second layer c-fermions  are very close to the ones of the bulk He$^3$\cite{vollhardt}; 
  the Coulomb terms $U_1$ and $U_2$  
 thus mainly re-normalize  the hybridization and hopping parameters, 
 inducing a dependence in the coverage  through 
  $ V = V_0 + V_1 n $  and $t=t_0 + t_1 n $, where
 n is the  total coverage in He$^3$.  
 
 Performing the slave boson decomposition  obtains the following Lagrangian \cite{comment2}:
 \bea \label{eqn2}
 {\cal L} & = &  \sum_{
\langle i, j \rangle, \sigma} \left [  f^\dagger_{i \sigma} 
\left ( ( \partial_\tau +E_0  + \lambda ) \delta_{ij} + b_i \alpha t b^\dagger_j  ) \right ) f_{j \sigma}   
\right  . \nonumber \\
  & + &  \left . c^\dagger_{i \sigma} \left (   ( \partial_\tau  - \mu  ) \delta_{ij}   + t  \right ) 
c_{j \sigma}     \right ]  \\
 & + &  V \sum_{i \sigma} \left   ( f^\dagger_{i \sigma} b_i c_{i \sigma} + h. c. \right )  
 +  J \left (  \sum_{
\langle i, j \rangle}  {\vec S}_i \cdot {\vec S}_j  - n_i n_j/4 \right ) \  , \nonumber \eea where 
${\vec S} = \sum_{\alpha \beta} f^\dagger_\alpha {\vec \sigma} f_ \beta $ is the spin operator
expressed in terms of the spinons only, with 
$ {\vec \sigma}$ the Pauli matrix and $\partial_\tau$ the partial derivative in imaginary time.
 $J = 2 (\alpha t)^2 / U $ is 
generated by a second order expansion of our model  in $U$. 
Alternatively, the $J$-term can be included ``ab initio'' in the model 
in consideration of the various ring exchange parameters
 generated \cite{roger2}. We assume that short range interactions stabilize a spin liquid with short range ferromagnetic character, in agreement with previous studies. 
 
  In order to  fit the experimental data, we  evaluate the dependance 
of  the bare parameters in coverage. Following \cite{tasaki}, we identify $E_{0}$ as the difference in the average potential energy between the first and the second layer. Here, the potential energy comes from the joined 
effects of  a) Van der Vaals potential  between the Graphite substrate and the layers 
$V_s (z) = \left ( 4 C_3^3 /  (27 D^2 )\right ) \ (1 /z^9  - C_3/ z^3) $  where $C_3 = 2092 K {\AA }^3$ is 
the  Van der Waals constant and $D  = 192 K $  \cite{roger2},    
and b) the Bernardes-Lennard Jones \cite{bernardes} potential acting between two 
He particles
$ V_{LJ} (r) = 4 \epsilon \left ( ( \sigma/ r )^{12} - (  \sigma/r )^6 \right )$ with $ \epsilon = 10.2 K$ and 
$ \sigma = 2.56 {\AA }$ is the hard core radius. Thus, $E_0= E_{l 1} - E _{l 2 } $ with 
$E_{l1 } = V_s (z_1) + \sum _i \rho_i \int r d r V_{LJ} (r ) $,  where $i$ indexes the contribution 
from the various layers(idem for $ E_{l 2}$).
We find  in Kelvin $E_{l1} = -10.495 -1.07 \ n  (K)$ while $E_{l2} = -8.73 -0.12 \ n  (K)$  with $n$ the {\it total }
coverage density in $n m^{-2}$ in good agreement with \cite{tasaki}.   
The parameter $E_0$ then reads $E_0 = -1. 79 - 0.95 \ n  (K)$.
The  half-bandwidths $D_f = 2 \alpha t = 2 \pi / m_f $   and $D_c = 2  t = 2 \pi / m_c$ are 
evaluated from \cite{pricaupenko} where the dependence in density is extracted from thermodynamic studies of the bulk. 
We find that the dependance in coverage is negligible compared to the one of $E_0$; $D_c = 1.1 K$, $D_f = 0.6 K $  
thus 
$ \alpha = 0.55 $. The relatively high value of $\alpha$ is to be contrasted with the typical 
values obtained in standard
Anderson lattice for rare earth compounds.  In this system the f-band is not particularly flat compared to the c-band.
  The parameter $J$ of the order of a few mK  can be extracted from the experiments \cite{saunders}. 
We take here $J= 4 mK$.
The main difficulty resides in evaluating the hybridization $V = V_0 + V_1 \ n $.   Noticing that the inter-layer spacing is twice 
smaller than the
distance between the intra-layer sites, it is reasonable to expect that  the hybridization is bigger 
than the c-  half bandwidth.   In this paper  we have adjusted the values of $V_0$ and $V_1$ to fit the experimental data. 
 We find $V= 13.05 - 1.2 \ n $ , so that at the QCP $V_{c}=1.89 K$ is larger than $D_c$,  in agreement with the above observation.  

\begin{figure}
\includegraphics[width=2.3 in, angle = -90]{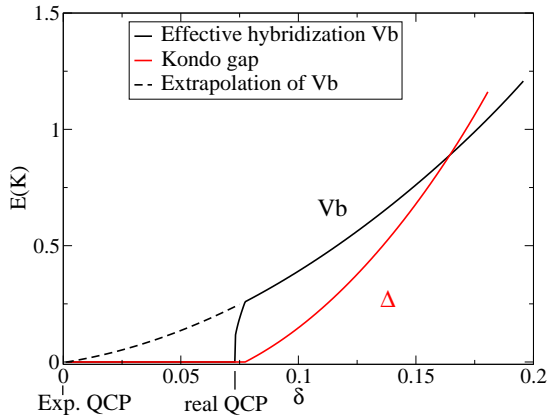}
\caption{ Mean-field phase diagram for the Anderson lattice model in $D=2$ applied to He$^3$ bi-layers.  
Following \cite{saunders} $\delta = 1- n/n_c$ with $n_c = 9.9 nm^{-2}$.
 The effective hybridization $Vb$    drops suddenly   at $\delta = 0.072$, indicating the real QCP. The experimental QCP is obtained
 by extrapolation of $ Vb$ to $T = 0$.  The Kondo gap  $\Delta $ (in red; color online) 
vanishes before the real QCP. }
\label{fig1}
\end{figure}

The mean-field equations are obtained by making a uniform and static approximation  for the holon
creation (annihilation)  operators $b^\dagger (b)$ in (\ref{eqn2}), then  evaluating the free energy $F$
and  solving for  the two equations 
$\partial F / \partial b = 0 $ and $\partial F/ \partial \lambda = 0 $.  
 We plotted in Fig.1 the effective hybridization
$b V$ as a function of the coverage (the unit for the coverage is identical to the 
experimental ones $ \delta  = ( n _c - n)/n_c $ with $ n_c = 9.9 nm ^{-2} $ ) and 
the ``Kondo gap'' $\Delta$ defined
as the energy difference between the  upper band and the Fermi energy.
The mean-field phase diagrams has two main features. First we observe an ``elbow'' in the order parameter  
as a function of the coverage, corresponding to the set up of the  Kondo phase (strong hybridization regime). The sharp change of behavior observed  corresponds to the emptying of the upper band. Note that the opening of 
 the Kondo gap occurs at the same point.
   The model gives an explanation for the mysterious 
observation that the field-induced magnetization (or static spin susceptibility) starts to grow
 before the experimental QCP is reached.
   In our model,  the point at which the magnetization starts to grow  is the  physical QCP, which differs from the one
obtained experimentally, which corresponds to the extrapolation of the linear regime to $T=0$. The static magnetic susceptibility is expected to  grow 
quickly in  as soon as the first layer localizes,  since the spin liquid parameter is small 
$J \sim 4  m K$. 
Moreover,  from our theory the Kondo gap  has to vanish  {\it before } 
 the QCP is reached. This comes from the observation that  at half filling (corresponding to the real QCP), the f-band is half filled,  hence constraining the bottom of the first layer to sit below the Fermi level.
This feature is observed experimentally, if we identify the Kondo gap as the activation 
gap extracted  from the thermodynamic 
measurements of \cite{saunders}.

 The mean-field value for the QCP reads $J/t = \exp{[ E_0  D_c/ V^2 ]}$ as in \cite{us}. 
We understand that any small variation in the value of $V$ with coverage has an exponential impact on the position of the QCP, justifying our option of adjusting the value of V to fit the data. 

\begin{figure}
\includegraphics[width= 2.9 in]{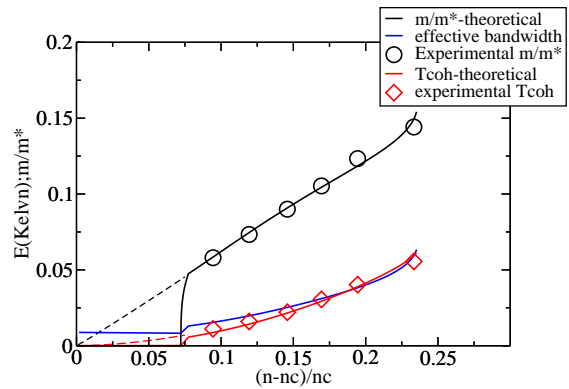}
\caption{ Coherence temperature (K) ,  inverse effective mass $ m/ m^*$ and 
effective spinon bandwidth $D_{eff}= \alpha^\prime D_c $ (K) 
in the Anderson lattice model for the
He$^3$ bi-layers. $ \alpha^\prime = b^2 \alpha + J / t $. The dots are experimental data from\cite{saunders}.  The effective bandwidth sets the upper temperature of the quantum critical regime.
The fitting parameters for this model are detailed in the text. }
\label{fig2}
\end{figure}

 We now turn to the fluctuations.  
  A striking observation  inferred from the experimental data, is the absence of  quantum critical (QC) regime  in temperature. A  Curie law for the spin susceptibility is observed  at very low temperatures in the  localized phase  and directly above $ T_{coh}$. Generically,  the upper energy scale of the QC fluctuations is determined by the first irrelevant operator.  In our model the formation of the spin liquid  and of the Fermi liquid are the two mechanisms  for quenching the entropy. Hence the temperature  $T^*$ below  which  the entropy $R ln 2 $ is quenched  goes like $Max[ D_{eff}, T_{coh}]$, with $D_{eff}$ the effective bandwidth of the spinons  and $T_{coh}$ the coherence temperature coming from the quantum fluctuations.  The dependence of $D_{eff}$ and $T_{coh}$ with coverage are depicted in Fig 2.We see that  $D_{eff}$ is  of the order of $J \sim 4 mK $ in the localized phase and follows $T_{coh}$ in the  Fermi liquid phase. The fact that $J$  in this system is a remarkably small energy scale, compared to heavy fermions, is thus the reason why  the QC regime is reduced to much lower temperatures.  Quantum fluctuations are thus most clearly seen through the variation of the effective mass and the shape of  $T_{coh}$.
 
 The quantum fluctuations  are  in the same 
 universality class as the ones
 of the Kondo breakdown model\cite{us}.
 The fluctuation spectrum
 in the intermediate energy regime  admits the dynamical exponent $z=3$,
 \beq \label{eqn3}
 D^{-1}_b ( q, \Omega_n) = D_0^{-1} \left [  q^2 + \xi^{-2}  +  \frac{\gamma |\Omega_n | }{ \alpha^\prime  q } \right ] \  ,  \eeq with $D_0 = 4 k_F^2 / (\rho_0  V^2 ) $, 
 $\xi$ is the correlation length associated with the fluctuations of b, $ \gamma =  m V^2 D_0 / ( \pi v_F ) $  $\alpha^\prime = b^2 \alpha + J/t$ and $ \rho_0 = m_c/  ( 2 \pi )$ is the
  c-fermions density of states. 
 The boson mass is given by $ m_b = D_0^{-1}  \ \xi^{-2}$ evaluated at $T=0$. We evaluate it  by differentiating twice the mean-field equations with 
 respect to the bosonic field $b$
 and  evaluating the result at the mean-field saddle point. For the effective mass, we use the  Luttinger-Ward expression of the free 
 energy typical of Heisenberg-type theories, analogous
 to the one derived in \cite{maslov-chub} $ F = F_{FG} +  T/2 \  \sum_n  \int d^2 q/ (2 \pi)^2  \log \left [ D^{-1} (q, \Omega_n ) \right ] $
  where $F_{FG}$ is the free-energy of
 the system at the mean-field. Since the system at the mean-field consists of two hybridized bands, one obtains after diagonalization
 \beq \label{eqn4}
 F = - \frac {\pi T^2}{6} \left [ 2 \pi ( \rho_1 + \rho_2 ) + \frac{ \gamma \xi}{ 4 \alpha^\prime} \right ] \ , \eeq where $\rho_1 (\rho_2)$ are the density of states of the
 upper(lower) bands.    The effective mass thus reads $ m^* =  2 \pi ( \rho_1 + \rho_2 )  + \gamma \xi / ( 4 \alpha^\prime ) $. 
 The coherence temperature is computed by  evaluating the corrections to scaling to the boson propagator (\ref{eqn3}); namely the one 
 loop diagrams responsible
 for the temperature dependence of $m_b (T)= D_0^{-1} \ \xi^{-2}(T) $.  Following\cite{jerome} we find (remember $z=3$ here ) 
 \beq \label{eqn5}
 m_b(T) = m_b (T=0)  + C   \  T Log T  \ , \eeq where $C$ had to be adjusted to $C = 7.5 \ 10^{-3}$ to fit the data, while the analytic evaluation gives $C = DJ/ (6 V_{c}^2)$. The coherence temperature $T_{coh}$ obtains
 when the  cross-over condition $m_b (T)= 0$
 is satisfied. The results for $m/m^*$ and $T_{coh}$ are presented in Fig.2 and directly compared to experiments.    Since we work within a slave boson saddle point approximation,  the need of one fitting parameter 
 for the amplitude of the coherence 
 temperature is to be expected.  
  The exponents can be understood in 
 a simple way.
 For $z=3$ theories in the Fermi liquid phase, the effective mass goes like the correlation length \cite{jerome}  $ m/ m^* \sim \xi^{-1}$. From the 
 dispersion of  the
  boson mode we see that $\xi^{-1} \sim \sqrt{m_b} \sim b $.  Now  the coherence temperature goes like $b^2$. In the regime where $b$ varies linearly with 
  the coverage
   $n$ we thus get
  \beq \label{eqn6}
  \begin{array}{ll}
  m/m^* \sim cst - n \ , \; \; \; \; & \; \; \; T_{coh} \sim  (cst - n )^2 \  . \end{array} \eeq   
    
   In conclusion we have performed a study of quantum criticality in  He$^3$  bi-layers by mapping the 
experiments on an extended version of  the Anderson lattice 
   in two dimensions.
   We examined the possibility for the Kondo breakdown QCP to  be responsible for the quantum fluctuations observed.    
The system 
   of He$^3$ bi-layers  enables
   to directly test the theory from  the bare  parameters. Our model is successful in that 
   \begin{itemize}
   \item  it explains the  occurrence of two QCPs  seemingly observed experimentally. Our interpretation 
is  that  one of the QCPs
   observed experimentally corresponds to  the extrapolation to $T=0$ of an intermediate energy regime;
   \item it predicts  that the activation gap extracted from thermodynamic measurements vanishes 
before the QCP is reached;
   
   \item it gives  exponents for the variation of the  effective mass $m/m^*$ and  the coherence 
temperature $T_{coh}$ in good agreement 
   with experiment, 
   so that a direct fitting of the data  is possible.
   \end{itemize}

We have used three fitting parameters to account for the 
prefactors of the effective mass 
   and coherence temperature as well as to precisely determine  the position 
of the QCP from the bare parameters.
   Our study is the first case where  an itinerant QCP  showing non Fermi liquid  behavior 
is used to 
   fit experimental data from the bare parameters.
      

 Useful discussions with H. Godfrin, G. Misguich, M. Neumann, J. Ny\'eki, O. Parcollet, M. Roger and J. Saunders are acknowledged. This work is supported by the French National Grant ANR26ECCEZZZ.

\end{document}